\documentclass[10pt,conference]{IEEEtran} % IEEEtran requires: pdflatex, bibtex8
\usepackage[english]{babel}
\usepackage{times}
\usepackage{caption}
\usepackage{graphicx}
\newcommand{\fig}[4][htbp]{% [ position ], options, file, caption
\begin{figure}[#1]
\centering%
\includegraphics[#2]{#3}
\caption[#4]{#4.}\label{fig:#3}
\end{figure}}
\usepackage{etoolbox}
\usepackage{siunitx}
\usepackage{booktabs}
\usepackage{color}

\usepackage[pdftex, pagebackref=false, colorlinks=true, linkcolor=black, 
citecolor=black, filecolor=black, urlcolor=black, plainpages=false]{hyperref}
\usepackage{url}% break long URL
\hypersetup{%
pdfauthor={Nuno A. C. Henriques},
pdftitle={SensAI+Expanse Emotional Valence Prediction Studies with Cognition 
and Memory Integration},
pdfkeywords={emotional valence prediction, context, cognition, memory, HAI},
pdfproducer={nunoachenriques.net}}

\usepackage{glossaries}
\setacronymstyle{long-short}
\newacronym{anr}{ANR}{Application Not Responding}
\newacronym{automl}{AutoML}{Automated Machine Learning}
\newacronym{gps}{GPS}{Global Positioning System}
\newacronym{hai}{HAI}{Human-Agent Interaction}
\newacronym{mgrs}{MGRS}{Military Grid Reference System}
\newacronym{shap}{SHAP}{SHapley Additive exPlanations}
\newacronym{weird}{WEIRD}{Western, Educated, Industrialized, Rich, and 
Democratic}
\newacronym{xai}{XAI}{Explainable Artificial Intelligence}

\def\rleft{$\mathsf{\left[10, 34\right)}$}
\def\rleftfemale{$\mathsf{\left[10, 34\right) female}$}
\def\rleftmale{$\mathsf{\left[10, 34\right) male}$}
\def\rright{$\mathsf{\left[34, 70\right)}$}
\def\rrightfemale{$\mathsf{\left[34, 70\right) female}$}
\def\rrightmale{$\mathsf{\left[34, 70\right) male}$}
\def\xgboostname{Extreme Gradient Boosting}

\def\nunoachenriques{\href{https://nunoachenriques.net/}{Nuno A. C. Henriques}}

\def\android{\href{https://developer.android.com/}{Android}}
\def\aposteriori{\textit{a posteriori}}
\def\pvalue{$p\ \textnormal{value}$}
\def\sensaigit{https://gitlab.com/nunoachenriques/sensei}
\def\sensai{\href{\sensaigit}{SensAI}}
\def\sensaiexpansegit{https://gitlab.com/nunoachenriques/sensei-expanse}
\def\sensaiexpanse{\href{\sensaiexpansegit}{SensAI Expanse}}
\def\expanse{\href{\sensaiexpansegit}{Expanse}}
\def\sensaiplusexpanse{\sensai+\expanse}
\def\twitter{\href{https://twitter.com/}{Twitter}}

\begin{document}
%%% IARIA enforced: https://www.iaria.org/editorialrules.html %%%%%%%%%%%%%%%%%%
\pagenumbering{gobble}
\title{\textbf{\Large \sensaiplusexpanse\\[-1.5ex] Emotional Valence Prediction 
Studies with Cognition and Memory Integration}\\[0.2ex]}
\author{\IEEEauthorblockN{~\\[-0.4ex]\large 
\nunoachenriques\\[0.3ex]\normalsize}
\IEEEauthorblockA{BioISI\\
Faculdade de Ci\^{e}ncias\\
Universidade de Lisboa\\
Portugal\\
{\tt nach@edu.ulisboa.pt}}
\and
\IEEEauthorblockN{~\\[-0.4ex]\large Helder Coelho\\[0.3ex]\normalsize}
\IEEEauthorblockA{BioISI\\
Faculdade de Ci\^{e}ncias\\
Universidade de Lisboa\\
Portugal\\
{\tt hcoelho@di.fc.ul.pt}}
\and
\IEEEauthorblockN{~\\[-0.4ex]\large Leonel Garcia-Marques\\[0.3ex]\normalsize}
\IEEEauthorblockA{CICPSI\\
Faculdade de Psicologia\\
Universidade de Lisboa\\
Portugal\\
{\tt garcia\_marques@sapo.pt}}}
\maketitle
\begin{abstract}
The humans are affective and cognitive beings relying on memories for their
individual and social identities. Also, human dyadic bonds require some common
beliefs, such as empathetic behaviour for better interaction. In this sense,
research studies involving human-agent interaction should resource on affect,
cognition, and memory integration. The developed artificial agent system
(\sensaiplusexpanse) includes machine learning algorithms, heuristics, and
memory as cognition aids towards emotional valence prediction on the interacting
human. Further, an adaptive empathy score is always present in order to engage
the human in a recognisable interaction outcome. The developed system encompass
a mobile device embodied interacting agent (\sensai) plus its Cloud-expanded
(\expanse) cognition and memory resources. The agent is resilient on collecting
data, adapts its cognitive processes to each human individual in a learning best
effort for proper contextualised prediction. The current study make use of an
achieved adaptive process. Also, the use of individual prediction models with
specific options of the learning algorithm and evaluation metric from a previous
research study. The accomplished solution includes a highly performant
prediction ability, an efficient energy use, and feature importance explanation
for predicted probabilities. Results of the present study show evidence of
significant emotional valence behaviour differences between some age ranges and
gender combinations. Therefore, this work contributes with an artificial
intelligent agent able to assist on cognitive science studies, as presented.
This ability is about affective disturbances by means of predicting human
emotional valence contextualised in space and time. Moreover, contributes with
learning processes and heuristics fit to the task including economy of cognition
and memory to cope with the environment. Finally, these contributions include an
achieved age and gender neutrality on predicting emotional valence states in
context and with very good performance for each individual.
\end{abstract}


%%%%%%%%%%%%%%%%%%%%%%%%%%%%%%%%%%%%%%%%%%%%%%%%%%%%%%%%%%%%%%%%%%%%%%%%%%%%%%%%
\section{Introduction}

The human agents may be seen like self-consciousness, emotion-driven, cognitive
beings with a bond between the evolutionary way of emotions and their supporting
physical structure, as proposed in \cite{Damasio2010EN}. In a sense, an agent's
behaviour depends on its affective states besides cognition to cope with the
environment. Thus, the ability to predict some dimension of those affects would
undoubtedly be of great value towards better adaptation and interaction with
others. Furthermore, an artificial agent, like in humans, may be conceived with
a decision process influenced by facts, reason, memories \cite{Wood2012} and
emotional pressure for the basic needs to the entity's homeostasis
\cite{Gadanho2001}. Additionally, using the concept of
empathy~\cite{Stueber2014}\cite{DeWaal2008} as a starting point for two-agent
bonding may bring better dyadic interaction and communication. Therefore, an
artificial agent adjusting empathetically towards the interacting human current
behaviour and affective state may be conceived
\cite{Picard1997}\cite{Perusquia-Hernandez2019}. Further, \gls{hai} should be
open to each entity own perception, such as the perceived human affective states
in a multimodal approach \cite{Tavabi2019}\cite{Tsiourti2018}. Accordingly, the
use of a wearable or mobile device, such as a smartphone seems suitable for the
task of collecting data towards affect sensing and reasoning. Twenty years ago,
the American College of Medical Informatics (ACMI) has already approached the
subject during the 1998 Scientific Symposium: ``Monitor the developments in
emerging wearable computers and sensors --- possibly even implantable ones ---
for their potential contribution to a personal health record and status
monitoring''~\cite{Greenes1998}. Currently, the mobile device as a sensing tool
for behavioural research is thriving with active discussions
\cite{Denecke2019}\cite{Felix2019} including exploration on correlates between
sensors' data and depressive symptom severity \cite{Saeb2015}.

This paper describes the \sensaiplusexpanse\ system ability to predict human
emotional valence states in geospatial and temporal context without evidence of
any bias regarding demographics, such as age and gender. This prediction ability
is supported by cognition and memory integration in learning mechanisms within
an adaptive process. Further, human population in current study comprise
distinct behaviour, age, gender, and place of origin. Meaning, the need for
proper adaptation on reasoning about each individual and respective collected
data in order to assure demographics neutrality. Accordingly, the system
developed for the affective studies encompasses an integration of cognition and
memory resources distributed between a smartphone embodied agent (\sensai) and
its Cloud-expanded (\expanse) continuity. \sensai\ collects data from multimode
sources, such as (a) device sensors (e.g., \gls{gps}); (b) current timestamp in
user calendar time zone; and (c) human text writings from in-application diary
and social network posts (e.g., \twitter). These written texts in (c) will be
subjected to a rule-based lexical processing \cite{Hutto2014} in order to obtain
a valence value, i.e., instantaneous sentiment analysis on demand with efficient
energy use. The human reports emotional valence ground truth by means of three
available buttons regarding positive, neutral, and negative sentiment classes.
Simultaneously, an empathy score progress bar is visible during this interaction
(Figure~\ref{fig:sensai---empathy-notification}) and may change on events, such
as the human reporting frequency. The score decays over time and it is designed
to be the human-agent current value of empathy. Complementary, the available
algorithms and heuristics included in \expanse\ are used by a pipeline process
in order to reason about the data set of each human individual. These machine
learning services comprise (a) unsupervised algorithms, such as location
clustering parameters auto discovery; and (b) supervised ones, such as learning
hyperparameters auto tuning. Regarding the learning model and evaluation metric,
the choice is based on a previous study \cite{Henriques2019} where \xgboostname\
and F1 score achieved good results for several data sets. As depicted in
Figure~\ref{fig:prediction---f1-xgboost-bin}, the prediction performance of the
agent achieved a score greater than $0.9$ for almost two thirds of the
population (31 eligible out of 49 from a total of 57), one third distributed
between $0.7$ (exclusive) and $0.9$, with only one subject scoring slightly less
than $0.7$ ($0.68$).

\fig{scale=1.0}{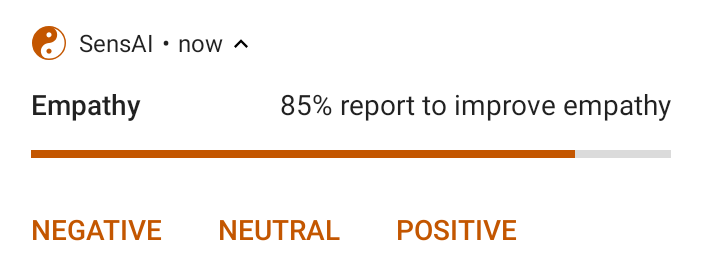}{Empathy score notification and 
emotional valence report buttons}

\fig{scale=0.4}{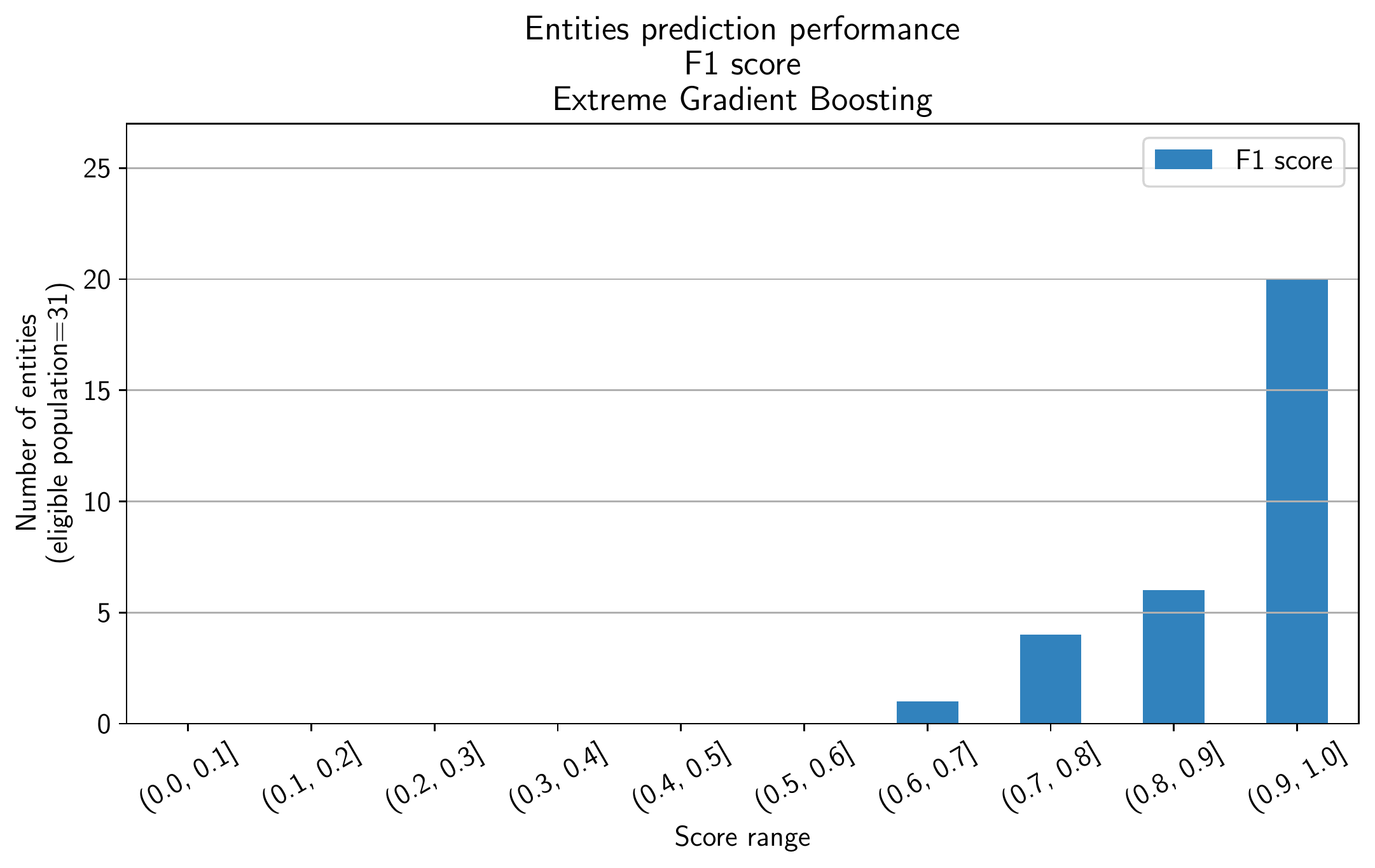}{Prediction performance on all 
individuals by F1 score range}

This first section introduced the current investigation purpose and the work
done so far. Next, Section \ref{sec:cognition} briefly describes the cognitive
and memory mechanisms in place for the developed system reasoning capabilities
on emotional valence prediction. Section \ref{sec:study} describes the research
study including the followed method and the achieved results. Section
\ref{sec:limitations} discloses the scope restrictions of this and similar
studies using smartphone sensing. Finally, Section \ref{sec:conclusion}
summarises the outcomes and presents a future perspective.

%%%%%%%%%%%%%%%%%%%%%%%%%%%%%%%%%%%%%%%%%%%%%%%%%%%%%%%%%%%%%%%%%%%%%%%%%%%%%%%%
\section{Cognition and Memory in \sensaiplusexpanse}\label{sec:cognition}

In order to enable \sensaiplusexpanse\ as a trustable research tool for
affective studies, this section briefly describes the achieved cognition and
memory integration in the developed system. In a sense, this agent is more than
an instrument by means of some bio-inspired concepts on its design and
implementation. Regarding natural brains, to some extent may be seen as mainly
about cognitive processing and memory use towards learning in order to adapt and
survive. All this, of course, besides identity, social and cultural aspects.
Moreover, emotions in humans are of great importance by helping on
decision-making and also with the persistence of episodic memories, amongst
other useful regulations \cite{Goertzel2010}\cite{Damasio1994EN}.
Simultaneously, ``Forgetting to remember''~\cite{Altmann2002} is required to
sustain adequate processing and storage when coping with the information stream
that flows through perception sensors. Thus, the concept of cognitive economy is
introduced comprising some exclusions from memory to foster savings by means of
(a) sustaining last collected data value without change from a sensor during a
well-defined time interval (e.g., 15 minutes); and (b) discarding data below
relevant thresholds (e.g., location changed less than 10 meters). The resulting
information gaps in (a) are filled \aposteriori\ in \expanse\ cognition. This
reconstruction is done strictly with the same time interval values used by
\sensai\ thus avoiding any data bias.

As already referred, adaptive mechanisms including cognitive and memory ones are
described in a previous paper \cite{Henriques2019}. Although, there are some of
this \gls{automl} process aspects relevant to emphasise once more. Regarding
system communication, \sensai\ and its \expanse\ are connected through an
end-to-end secure Web service. The data flows mainly from the mobile device
sensors collected by \sensai\ to the \expanse\ storage for posterior processing
in a pipeline. \sensai\ has in-device reasoning and memory abilities yet the
\expanse\ does the heavy work by means of several adaptive mechanisms regarding
collected data from human behaviour. In the end, the system functions as a
distributed, fault-tolerant, mobile and Cloud-based artificial intelligent
agent. It may be used as a robust, continuously, online research tool for
gathering and processing human emotional valence data towards contextualised
predictions, and also affective studies.

\subsection{\sensai}

The mobile device agent is embodied, encompassing several functions as a whole, 
towards proper interaction, data collecting, reasoning, and storage. The ground 
truth values for emotional valence prediction are reported by humans. The main 
interface includes three emoticon in buttons representing the available 
discrete valence classes of negative, neutral, and positive, as in the 
persistent notification (Figure~\ref{fig:sensai---empathy-notification}). 
Further, this mechanism is robust to interaction bias, such as high-frequency 
repeated button (emoticon) clicks. Also, to cases of mistaken valence reported 
and promptly corrected by an additional hit on a different class. Moreover, a 
simple yet effective heuristic of accounting only for the last hit during a 
defined short time interval is in place. All these happenings are properly 
contextualised by collecting the location and moment of the event. \sensai\ is 
focused mainly on:
\begin{LaTeXdescription}
\item[Smooth \gls{hai}] towards engaged empathetic relationship, reasoning about
current emotional valence state of the human, and keeping all interaction very
much passive only with seldom actions. The empathetic engagement relies on a
score to be perceived by the human as the \gls{hai} empathy level. This metric
is sensible to the frequency of human reporting. It decays over time (e.g.,
24-hour cycle) and the decay rate may change with other actions, such as pausing
the data collection. The agent writes some periodic messages using the
in-application diary with useful information, such as the detected activity
(e.g., running) and the current computed emotional valence value. All these
actions, specifically empathy level notifications, are kept silent, only
notifying when the human is interacting. Additionally, displays summary data
such as physical activity by means of six recognised classes (e.g., walking) in
a main dashboard. Also, a sentiment chart is available with interactive zoom and
pan along the local memory chronological limit (e.g., 28 days) of sentiments
reported and detected (e.g., Twitter status).
    
\item[Efficient data collection] as much as possible and allowed in a relatively
low energy consumption. Accordingly, a practical data acquisition rhythm, such
as $active=2s$, $inactive=8s$, $f=1/5Hz$, and $D=20\%$ is devised and
implemented in order to acquire relevant data without too much power drain
(below two-digit percentage points on average for several assessments). This
rhythm and other thresholds (e.g., $100ms$ minimum interval between sensor
fetches) may be subjected to automatic adaptation within environment changes.
Regarding the sentiment analysis \cite{Hutto2014} of written texts, a custom
heuristic is implemented \cite{Henriques2019} integrating language detection,
translation, and emoticon processing. All in a best effort to get the sentiment
value along with the language (English and Portuguese supported). This process
adapts to the cases of mixed languages, emoticon-only text, and no language
detected but emoticon available to extraction on analysing these short messages.
Additionally, a display with the statistics about all sensors collected events
is available. It includes sensor event count in local memory and already Cloud
synced, last data sync, and percentage of collecting activity relative to
application existence.

\item[Environment adaptation] in order to keep local resources healthy and
survive to sudden contingencies, such as application crashes. An
homeostasis-based implementation runs periodic checks, such as database health
and data feed. It will take proper actions, such as recovering the data
collection from sensors, and even a local database maintenance. The agent does a
system registration at first start to deal with device boot and application
upgrade special states. Also, guarantees the reviving by the \android\ operating
system in cases of unexpected crash and removal from the running state.
Regarding Cloud data syncing, it is robust to failures using a mechanism
inspired in the relational database management system transaction. Only synced
data will be marked for removal after local cache persistence threshold.
Moreover, if no suitable data connection is available then it will adapt by
increasing verification frequency for further try to sync. All these mechanisms
of local cache and Cloud sync are paramount to keep healthy memory consumption
and guarantee proper data collection.
\end{LaTeXdescription}

\subsection{\sensaiexpanse}

The \expanse\ comprises Cloud-expanded resources for the \sensai\ agent. These
are able to supplement the smartphone restricted local memory, processing, and
power. In a sense, it augments the agent cognition and memory. Regarding
persistence, there is storage available for data since first value collected
until the present. All data is secured and stored anonymously. Because of an
efficient data collection, fewer data stored represents more after proper
processing. There is a step during the machine learning pipeline where a
transformation acts on cleaning and reconstructing collected data. This includes
upsampling data within the actual thresholds previously used to save resources
in the mobile device. Moreover, processing all eligible data through the
available myriad of heuristics and other algorithms towards \gls{automl}
requires (a) an adaptation to the multiple human behaviours revealed in the data
set; and (b) Bayesian efficient auto discovery on parameters.

The developed custom pipeline for \sensai\ learning uses various heuristics and
other algorithms towards \gls{automl}. These include data classes (negative,
neutral, positive) imbalance (reports count) degree check from \cite{Zhu2018}.
And also, a custom heuristic for class verification and learning process
adaptation in cases, such as no reports for one (or even two) classes. After all
these reasoning, the achieved entities are the ones eligible for the machine
learning final step towards emotional valence prediction in context. Regarding
location, the learn process make use of the unsupervised HDBSCAN algorithm
clustering coordinates which accurately drops outliers. For each individual
case, the relevant steps may be summarised as (a) calling HDBSCAN on provided
\texttt{min\_samples} (using $[1, 10, 100]$) in order to find the best
\texttt{min\_cluster\_size} parameter; and before each call to the elected
supervised multi-class classification algorithm \xgboostname\ (b) an auto search
is run for the best cross validation $K$ splits (K-fold) regarding the algorithm
minimum count of accepted classes. Next, Bayesian optimisation is used for
hyperparameter auto tuning with cross validation for each specific model.
Finally, the model fit for each human current data is achieved and performance
metrics are computed. The actual knowledge from the learning process is stored
efficiently with a very small memory footprint. The system prediction ability is
ready to serve answers about contextualised emotional valence for each
individual.

%%%%%%%%%%%%%%%%%%%%%%%%%%%%%%%%%%%%%%%%%%%%%%%%%%%%%%%%%%%%%%%%%%%%%%%%%%%%%%%%
\section{Study}\label{sec:study}

This section encompasses the outcomes and method used in a research study 
running on a long-term interaction between \sensai\ and a population of human 
individuals in the field (anonymous participants outside the laboratory).

\subsection{Method}

The participants in this study are neither targeted nor recruited from anywhere.
Instead, the goal to avoid a laboratory known \cite{Henrich2010} frequent bias
of sampling only from \gls{weird} societies is accomplished by collecting data
in the wild and worldwide. Smartphone sensing \cite{Cornet2018} is in place  by
means of an \android\ application. Moreover, privacy is enforced by removing all
identifiable data before storing in \expanse. Furthermore, text in the Diary
activity is kept local and will be destroyed by \sensai\ uninstallation. The
user is informed on first install and by using the help option. A total of 57
participants (ten countries and four continents) installed \sensai, eight were
discarded for not sharing age and gender thus 49 remained eligible for analysis
before machine learning pipeline further restrictions. Moreover, some ten-year
age range classes revealed to be under-represented hence a dichotomy approach is
in place using age median ($M=34$). Therefore, a reasonable distribution of all
genders by the two age ranges, \rleft\ vs. \rright, is attained despite some
gender disproportion in each range. The age range count relative difference from
\rleft\ to \rright\ is only $4.2\%$ (one individual) as presented in
Table~\ref{tab:populationagegender}.

\captionsetup{font={footnotesize,sc},justification=centering,labelsep=period}%
\begin{table}[htbp]
\caption{Eligible population for analysis}
\label{tab:populationagegender}
\centering%
\begin{tabular}{lS[table-format=2]S[table-format=2]}
\toprule
{} & \multicolumn{2}{c}{Age range}\\
{} & {\rleft} & {\rright}\\
\midrule
Female &  9 &  9\\
Male   & 15 & 16\\
\midrule
Total  & 24 & 25\\
\bottomrule
\end{tabular}
\end{table}
\captionsetup{font={footnotesize,rm},justification=centering,labelsep=period}%

\subsection{Results}

Regarding the emotional valence states reported by the humans, there is evidence
of valence proportion differences between some groups. In order to assess the
statistical significance of this evidence depicted in
Figure~\ref{fig:analysis---behaviour-age-gender-percentage}, the Mann-Whitney U
test is used. The results presented in Table~\ref{tab:utestbehaviour} show that
the null hypothesis ($H_0$: two sets of measurements are drawn from the same
distribution) can be rejected for all but two tests, i.e., evidence of
significant differences on three comparisons of two groups each as described
next:

\begin{LaTeXdescription}
\item[\rleft\ vs. \rright] show differences between two age ranges mainly driven
by the female gender.

\item[\rleftfemale\ vs. \rrightfemale] evidence a difference in behaviour where
the older age group reported significantly less negative and neutral emotional
valence and an order of magnitude more positive reports.

\item[\rrightfemale\ vs. \rrightmale] evidence the overwhelming positive reports
by female versus male in this age range group.
\end{LaTeXdescription}  

\fig{scale=0.4}{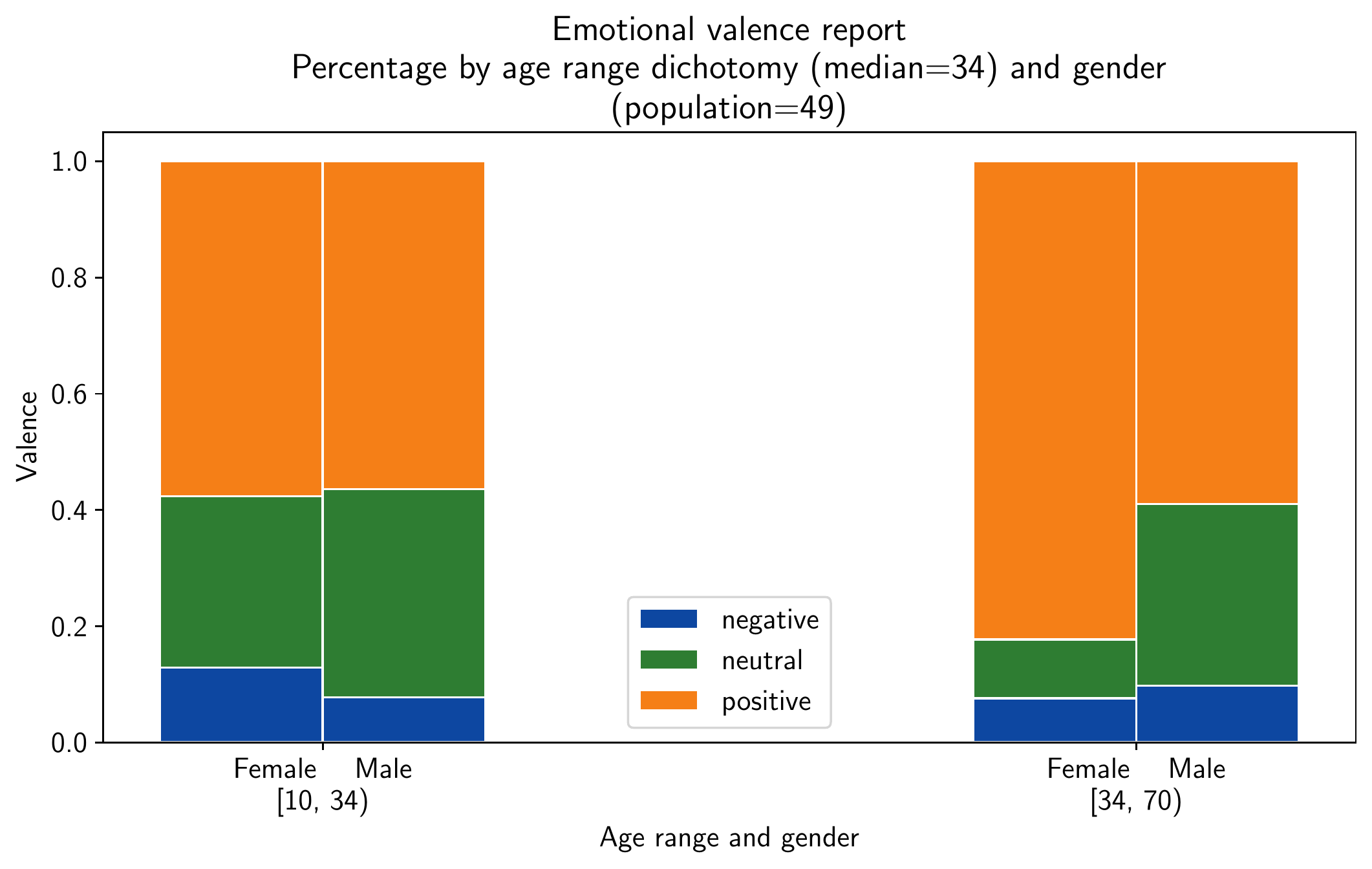}{Emotional valence 
reports percentage by age range and gender}

\captionsetup{font={footnotesize,sc},justification=centering,labelsep=period}%
\begin{table}[htbp]
\caption{Age and gender groups comparison 
(Figure~\ref{fig:analysis---behaviour-age-gender-percentage}): Mann-Whitney U 
test results}
\label{tab:utestbehaviour}
\centering%
\begin{tabular}{lS[table-format=1.3,table-figures-exponent=2,table-sign-exponent]c}
\toprule
Age range and gender & \pvalue & $H_0$ ($\alpha=0.05$)\\
\midrule
\rleft\ vs. \rright             & 1.161e-30 & rejected\\
\rleftfemale\ vs. \rrightfemale & 5.539e-14 & rejected\\
\rleftmale\ vs. \rrightmale     & 1.561e-01 & not rejected\\
\rleftfemale\ vs. \rleftmale    & 3.938e-01 & not rejected\\
\rrightfemale\ vs. \rrightmale  & 7.027e-67 & rejected\\
\bottomrule
\end{tabular}
\end{table}
\captionsetup{font={footnotesize,rm},justification=centering,labelsep=period}%

% PEER REVIEW change \cite{XXXXX} to \cite{Henriques2019}

Regarding the system prediction performance, a previous study
\cite{Henriques2019} revealed \xgboostname\ with F1 score as the best option on
average for the population individuals, as already depicted in
Figure~\ref{fig:prediction---f1-xgboost-bin}. The sample is reduced to 31
eligible individuals due to valence classes imbalance restrictions further
applied in the machine learning pipeline process. Regarding these entities, the
high scores achieved were independent of age and gender, no pattern whatsoever
was revealed. Further, inspection of the feature importance contribution to the
prediction revealed that most of the population is more sensible to the time
dimension than the location. Specifically, weekday (\texttt{moment\_dow}) is the
most, $64.5\%$ of the cases, influential for emotional valence predictions
followed by hour (\texttt{moment\_hour}) with $25.8\%$, and location
(\texttt{mgrs\_...}) with $9.7\%$. This ranking is obtained using \gls{shap} on
all entities prediction model for the feature ranked first on each entity.
Meaning that for almost all entities the moment, weekday and hour, is decisive
to obtain a prediction whereas for a few (e.g., entity 5 as depicted in
Figure~\ref{fig:prediction---e5-feature-overall-influence}) some locations
strongly compete for emotional valence prediction.

\fig{scale=0.4}{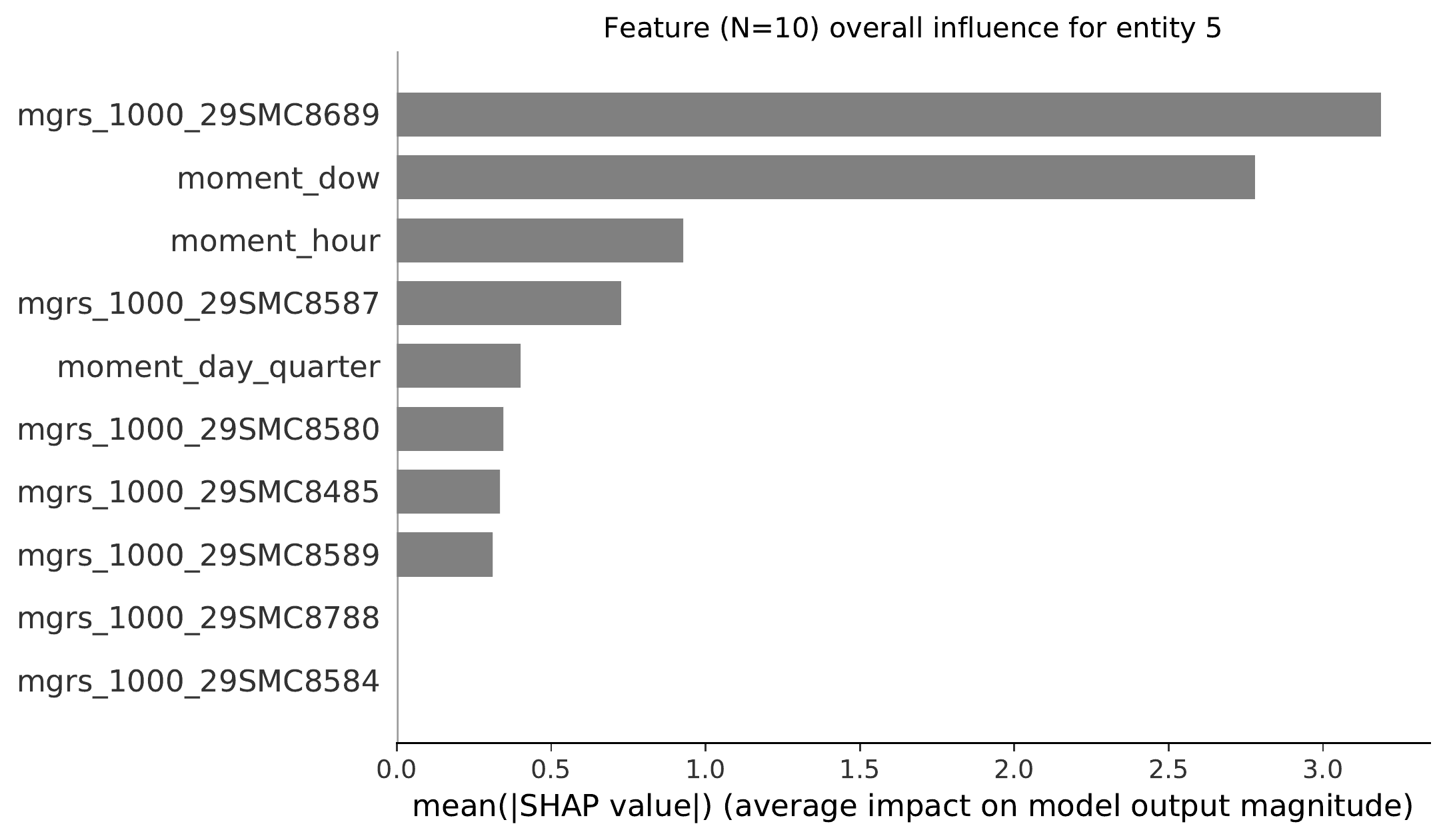}{Feature overall 
influence for entity 5}

Regarding the common behaviour, entity 24 is a typical individual where the time
dimension prevails on emotional valence prediction. In this case, Sunday is a
weekday where all locations are expected to have positive value predictions.
Accordingly, Figure~\ref{fig:prediction---e24-sunday-0800-map} shows the
expected positive emotional valence on a Sunday in the future at 8:00 a.m. for
the top seven locations (most influential).

\fig{scale=0.44}{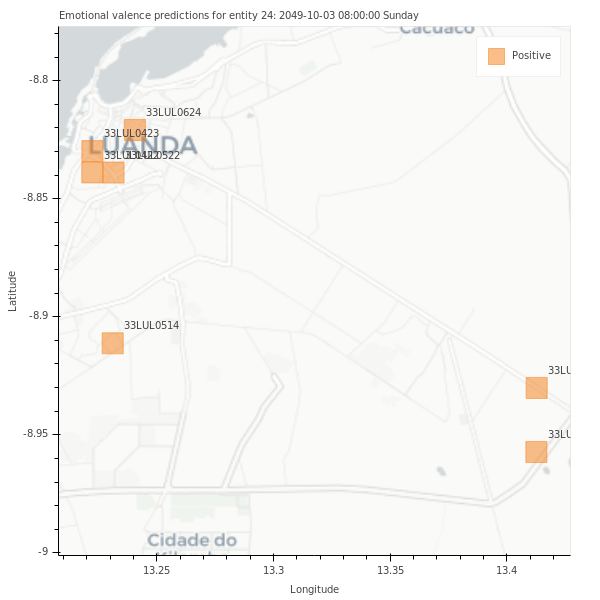}{Entity 24: Sunday valence 
probability for the top 7 locations}

Regarding the same top seven locations and hour of the day,
Figure~\ref{fig:prediction---e24-tuesday-0800-map} depicts a quite different day
from the previous Sunday. The computed probabilities are in order with the
expected for a business day for this entity at 8:00 a.m.

\fig{scale=0.44}{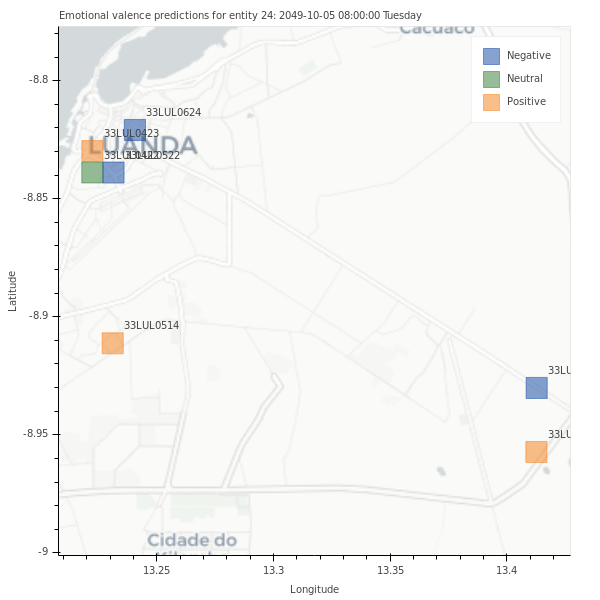}{Entity 24: Tuesday valence 
probability for the top 7 locations}

The map areas of emotional valence prediction measure \SI{1000}{\meter} square
side due to reasonable cell size for same place sentiment and a feasible number
of cells able to use as features in the learning process. Moreover, world map is
divided into cells following the \gls{mgrs}. These maps are obtained with a 
developed prototype (Jupyter notebook and Python) for prediction analysis 
online using the last \gls{automl} results for each entity. The tool is 
interactive including zoom and more information on hover each location.

%%%%%%%%%%%%%%%%%%%%%%%%%%%%%%%%%%%%%%%%%%%%%%%%%%%%%%%%%%%%%%%%%%%%%%%%%%%%%%%%
\section{Limitations}\label{sec:limitations}

Every scientific study has limitations. In order to clarify under which
conditions the results should be interpreted, the identified limitations of the
present work are (a) no prior health information about the users that may
impact the engagement effect including prediction result bias; (b) interacting
with a non-anthropomorphic versus human-like agent \cite{Schindler2016} may 
impact the emotional valence state reported; and (c) affective reactivity and
regulation gender differences on emotional response to context are not
considered as proposed in~\cite{McRae2008}. Although, gender and age neutrality
is achieved by \sensaiplusexpanse\ results on predicting emotional valence
states. Moreover, there is no evidence of any bias in the prediction scores 
achieved regarding the individual gender.

%%%%%%%%%%%%%%%%%%%%%%%%%%%%%%%%%%%%%%%%%%%%%%%%%%%%%%%%%%%%%%%%%%%%%%%%%%%%%%%%
\section{Conclusion and Future Work}\label{sec:conclusion}

This paper described the \sensaiplusexpanse\ system ability to predict emotional
valence states (a) in spatial and temporal context; (b) with very good
performance; and (c) age and gender neutral on revealing some individuals'
idiosyncrasies. Moreover, this smartphone sensing-based system is robust to
unexpected behaviours from humans, Cloud, and mobile demanding environments. The
\sensai\ agent first adapts to the operating system restrictions on mobile
resources use, such as keeping battery consumption below two-digit percentage
points on average. Then, it uses a myriad of heuristics and other algorithms in
order to achieve the best possible prediction whichever human behaviour
encounters within the collected data. The outcomes presented show evidence,
restricted to population and data samples in this study, of differences in
behaviour amongst some combinations of age ranges versus gender. Regardless,
\sensaiplusexpanse\ was able to adapt and learn to predict emotional valence
states with very good scores for every individual on average
(Figure~\ref{fig:prediction---f1-xgboost-bin}). Thus, \sensai\ is able to reveal
idiosyncratic factors on human's emotional valence changes without any bias
regarding age and gender. Moreover, adding features to the learning process may
reveal distinct factors not yet discovered, such as influence of physical
activity (e.g., riding a bike). The accelerometer data may be used to correlate
physical activity with valence (positive) state \cite{Lathia2017}. This course
of action may be taken as future work. Therefore, \sensaiplusexpanse\
contributes as a novel platform for affective and cognitive studies about human
emotional valence changes in context. Further, it may complement and eventually
supersede laboratory usually long-list self-appraisal questionnaires. Moreover,
it reinforces smartphone sensing contribution as a tool for personalised health
studies, such as emotional disturbances accompanied by healthcare professionals.
Furthermore, all the source code is published as free software under the Apache
License 2.0. Future work may include prior health information from each human.
Thus, adapting interaction and learning process towards better predictions.
Furthermore, \sensai\ may enable classifying options (e.g., Likert scale) at
some specific events for the human to grade the agent behaviour hence tailoring
future actions. This course of action may diminish the identified limitations
described in the previous section and contribute to better studies.

%%%%%%%%%%%%%%%%%%%%%%%%%%%%%%%%%%%%%%%%%%%%%%%%%%%%%%%%%%%%%%%%%%%%%%%%%%%%%%%%
\section*{Acknowledgement}

N.A.C.H. thanks Jorge M. C. Gomes for the precious contributions. This work is
partially supported by \textit{Universidade de Lisboa} [PhD support grant
May 2016--April 2019]. Partially supported by
\textit{Funda\c{c}\~{a}o para a Ci\^{e}ncia e Tecnologia} [UID/MULTI/04046/2019
Research Unit grant from FCT, Portugal (to BioISI)]. This work used the European
Grid Infrastructure (EGI) with the support of NCG-INGRID-PT.

%%%%%%%%%%%%%%%%%%%%%%%%%%%%%%%%%%%%%%%%%%%%%%%%%%%%%%%%%%%%%%%%%%%%%%%%%%%%%%%%
\bibliographystyle{IEEEtranNACH}\bibliography{sensai-cognitive}

\begin{thebibliography}{10}
\providecommand{\url}[1]{#1}
\csname url@samestyle\endcsname
\providecommand{\newblock}{\relax}
\providecommand{\bibinfo}[2]{#2}
\providecommand{\BIBentrySTDinterwordspacing}{\spaceskip=0pt\relax}
\providecommand{\BIBentryALTinterwordstretchfactor}{4}
\providecommand{\BIBentryALTinterwordspacing}{\spaceskip=\fontdimen2\font plus
\BIBentryALTinterwordstretchfactor\fontdimen3\font minus
  \fontdimen4\font\relax}
\providecommand{\BIBforeignlanguage}[2]{{%
\expandafter\ifx\csname l@#1\endcsname\relax
\typeout{** WARNING: IEEEtran.bst: No hyphenation pattern has been}%
\typeout{** loaded for the language `#1'. Using the pattern for}%
\typeout{** the default language instead.}%
\else
\language=\csname l@#1\endcsname
\fi
#2}}
\providecommand{\BIBdecl}{\relax}
\BIBdecl

\bibitem{Damasio2010EN}
A.~Dam{\'{a}}sio, \emph{{Self Comes to Mind: Constructing the Conscious
  Brain}}, 1st~ed.\hskip 1em plus 0.5em minus 0.4em\relax [``O Livro da
  Consci{\^{e}}ncia: A Constru{\c{c}}{\~{a}}o do C{\'{e}}rebro Consciente'',
  C{\'{i}}rculo de Leitores, ISBN:9789896441203], 2010.

\bibitem{Wood2012}
R.~Wood, P.~Baxter, and T.~Belpaeme, ``{A review of long-term memory in natural
  and synthetic systems},'' \emph{Adaptive Behavior}, vol.~20, no.~2, pp.
  81--103,  2012. DOI:\url{https://doi.org/10.1177/1059712311421219}

\bibitem{Gadanho2001}
S.~C. Gadanho and J.~Hallam, ``{Robot learning driven by emotions},''
  \emph{Adaptive Behavior}, vol.~9, no.~1, pp. 42--64,  2001.
  DOI:\url{https://doi.org/10.1177/105971230200900102}

\bibitem{Stueber2014}
\BIBentryALTinterwordspacing
K.~Stueber, \emph{{Empathy}}, winter2014~ed., E.~N. Zalta, Ed.\hskip 1em plus
  0.5em minus 0.4em\relax  Metaphysics Research Lab, Stanford University, 2014.
  URL:\url{http://plato.stanford.edu/archives/win2014/entries/empathy}
\BIBentrySTDinterwordspacing

\bibitem{DeWaal2008}
F.~B.~M. de~Waal, ``{Putting the altruism back into altruism: The evolution of
  empathy},'' \emph{Annual Review of Psychology}, vol.~59, pp. 279--300,  2008.
  DOI:\url{https://doi.org/10.1146/annurev.psych.59.103006.093625}

\bibitem{Picard1997}
R.~W. Picard, \emph{{Affective Computing}}.\hskip 1em plus 0.5em minus
  0.4em\relax Cambridge, MA, USA: MIT Press, 1997. ISBN:0-262-16170-2

\bibitem{Perusquia-Hernandez2019}
M.~Perusqu{\'{i}}a-Hern{\'{a}}ndez, D.~A.~G. J{\'{a}}uregui, M.~Cuberos-Balda,
  and D.~Paez-Granados, ``{Robot mirroring: A framework for self-tracking
  feedback through empathy with an artificial agent representing the self},''
  2019. arXiv:\url{http://arxiv.org/abs/1903.08524v1}

\bibitem{Tavabi2019}
L.~Tavabi, ``{Multimodal Machine Learning for Interactive Mental Health
  Therapy},'' in \emph{2019 International Conference on Multimodal Interaction
  on - ICMI '19}.\hskip 1em plus 0.5em minus 0.4em\relax  New York, New York,
  USA: ACM Press, 2019. DOI:\url{https://doi.org/10.1145/3340555.3356095} pp.
  453--456.

\bibitem{Tsiourti2018}
C.~Tsiourti, ``{Artificial agents as social companions: design guidelines for
  emotional interactions},'' phdthesis, Universit{\'{e}} de Gen{\`{e}}ve, 2018.

\bibitem{Greenes1998}
R.~A. Greenes and N.~M. Lorenzi, ``{Audacious Goals for Health and Biomedical
  Informatics in the New Millennium},'' \emph{Journal of the American Medical
  Informatics Association}, vol.~5, no.~5, pp. 395--400,  1998.
  DOI:\url{https://doi.org/10.1136/jamia.1998.0050395}

\bibitem{Denecke2019}
K.~Denecke, E.~Gabarron, R.~Grainger, S.~T. Konstantinidis, A.~Lau,
  O.~Rivera-Romero, T.~Miron-Shatz, and M.~Merolli, ``{Artificial Intelligence
  for Participatory Health: Applications, Impact, and Future Implications},''
  \emph{Yearbook of Medical Informatics},  2019.
  DOI:\url{https://doi.org/10.1055/s-0039-1677902}

\bibitem{Felix2019}
I.~R. Felix, L.~A. Castro, L.-f. Rodriguez, and O.~Banos, ``{Mobile sensing for
  behavioral research: A component-based approach for rapid deployment of
  sensing campaigns},'' \emph{International Journal of Distributed Sensor
  Networks}, vol.~15, no.~9, pp. 1--17,  2019.
  DOI:\url{https://doi.org/10.1177/1550147719874186}

\bibitem{Saeb2015}
S.~Saeb, M.~Zhang, C.~J. Karr, S.~M. Schueller, M.~E. Corden, K.~P. Kording,
  and D.~C. Mohr, ``{Mobile Phone Sensor Correlates of Depressive Symptom
  Severity in Daily-Life Behavior: An Exploratory Study.}'' \emph{Journal of
  medical Internet research}, vol.~17, no.~7, p. e175,  2015.
  DOI:\url{https://doi.org/10.2196/jmir.4273}

\bibitem{Hutto2014}
C.~J. Hutto and E.~Gilbert, ``{VADER: A Parsimonious Rule-based Model for
  Sentiment Analysis of Social Media Text},'' in \emph{Proceedings of the
  Eighth International AAAI Conference on Weblogs and Social Media}, Ann Arbor,
  Michigan, USA, 2014, pp. 216--225.

\bibitem{Henriques2019}
N.~A.~C. Henriques, H.~Coelho, and L.~Garcia-Marques, ``{SensAI+Expanse
  Adaptation on Human Behaviour Towards Emotional Valence Prediction},'' pp.
  1--6,  2019. arXiv:\url{http://arxiv.org/abs/1912.10084v4}

\bibitem{Goertzel2010}
B.~Goertzel, R.~Lian, I.~Arel, H.~de~Garis, and S.~Chen, ``{A world survey of
  artificial brain projects, Part II: Biologically inspired cognitive
  architectures},'' \emph{Neurocomputing}, vol.~74, no. 1-3, pp. 30--49,  2010.
  DOI:\url{https://doi.org/10.1016/j.neucom.2010.08.012}

\bibitem{Damasio1994EN}
A.~Dam{\'{a}}sio, \emph{{Descartes' Error: Emotion, Reason, and the Human
  Brain}}, 2nd~ed.\hskip 1em plus 0.5em minus 0.4em\relax [``O Erro de
  Descartes: Emo{\c{c}}{\~{a}}o, Raz{\~{a}}o e o C{\'{e}}rebro Humano'',
  C{\'{i}}rculo de Leitores, ISBN:9724211290], 1994.

\bibitem{Altmann2002}
E.~M. Altmann and W.~D. Gray, ``{Forgetting to remember: The functional
  relationship of decay and interference},'' \emph{Psychological Science},
  vol.~13, no.~1, pp. 27--33,  2002.
  DOI:\url{https://doi.org/10.1111/1467-9280.00405}

\bibitem{Zhu2018}
R.~Zhu, Z.~Wang, Z.~Ma, G.~Wang, and J.~H. Xue, ``{LRID: A new metric of
  multi-class imbalance degree based on likelihood-ratio test},'' \emph{Pattern
  Recognition Letters}, vol. 116, pp. 36--42,  2018.
  DOI:\url{https://doi.org/10.1016/j.patrec.2018.09.012}

\bibitem{Henrich2010}
J.~Henrich, S.~J. Heine, and A.~Norenzayan, ``{The weirdest people in the
  world?}'' \emph{Behavioral and Brain Sciences}, vol.~33, no. 2-3, pp. 61--83,
   2010. DOI:\url{https://doi.org/10.1017/S0140525X0999152X}

\bibitem{Cornet2018}
V.~P. Cornet and R.~J. Holden, ``{Systematic review of smartphone-based passive
  sensing for health and wellbeing},'' \emph{Journal of Biomedical
  Informatics}, vol.~77, pp. 120--132,  2018.
  DOI:\url{https://doi.org/10.1016/j.jbi.2017.12.008}

\bibitem{Schindler2016}
S.~Schindler and J.~Kissler, ``{People matter: Perceived sender identity
  modulates cerebral processing of socio-emotional language feedback},''
  \emph{NeuroImage},  2016.
  DOI:\url{https://doi.org/10.1016/j.neuroimage.2016.03.052}

\bibitem{McRae2008}
K.~McRae, K.~N. Ochsner, I.~B. Mauss, J.~J.~D. Gabrieli, and J.~J. Gross,
  ``{Gender Differences in Emotion Regulation: An fMRI Study of Cognitive
  Reappraisal},'' \emph{Group Processes {\&} Intergroup Relations}, vol.~11,
  no.~2, pp. 143--162,  2008.
  DOI:\url{https://doi.org/10.1177/1368430207088035}

\bibitem{Lathia2017}
N.~Lathia, G.~M. Sandstrom, C.~Mascolo, and P.~J. Rentfrow, ``{Happier People
  Live More Active Lives: Using Smartphones to Link Happiness and Physical
  Activity},'' \emph{PLOS ONE}, vol.~12, no.~1, p. e0160589,  2017.
  DOI:\url{https://doi.org/10.1371/journal.pone.0160589}

\end{thebibliography}
\end{document}